\documentclass{aa}  

\usepackage{graphicx}
\usepackage{tabularx}
\usepackage{txfonts}
\usepackage[colorlinks=true,citecolor=blue,linkcolor=blue]{hyperref}
\usepackage{soul}

\newcommand{\ho}{$H_0$}
\newcommand{\zd}{$z_d$}
\newcommand{\zs}{$z_s$}

\newcommand{\hunit}{km s$^{-1}$ Mpc$^{-1}$}
\newcommand{\vdunit}{km s$^{-1}$}
\newcommand{\as}{$^{\prime \prime}$}
\newcommand{\referee}[1]{\textcolor{black}{{#1}}}   %{\textbf{{#1}}} %{\textcolor{black}{{#1}}}
\newcommand{\newadd}[1]{\textcolor{black}{{#1}}}

\begin{document} 

\title{TDCOSMO}

\subtitle{XI. New lensing galaxy redshift and velocity dispersion measurements from Keck spectroscopy of eight lensed quasar systems}

\author{P. Mozumdar, \inst{1}
        C. D. Fassnacht, \inst{1}
        T. Treu,\inst{2}
        C. Spiniello, \inst{3,4}
        and
        A.~J.~Shajib\inst{5, 6}\fnmsep\thanks{NFHP Einstein Fellow}
       %\and
       %\fnmsep\thanks{Just to show the usage
      %of the elements in the author field}
       }

\institute{Department of Physics and Astronomy, University of California, Davis, CA 95616, USA \\
          \email{pmozumdar@ucdavis.edu}
    \and
		  Department of Physics and Astronomy, University of California, Los Angeles, CA 90095, USA
     \and
      Sub-Dep. of Astrophysics, Dep. of Physics, University of Oxford, Denys Wilkinson Building, Keble Road, Oxford, OX1 3RH, UK
      \and
      INAF -  Osservatorio Astronomico di Capodimonte, Via Moiariello  16, 80131, Naples, Italy
     \and
     Department  of  Astronomy  \&  Astrophysics,  University  of Chicago, Chicago, IL 60637, USA
     \and
     Kavli Institute for Cosmological Physics, University of Chicago, Chicago, IL 60637, USA
     %    University of Alexandria, Department of Geography, ...\\
     %    \email{c.ptolemy@hipparch.uheaven.space}
     %    \thanks{The university of heaven temporarily does not
    %             accept e-mails}
         }

%\date{Received September 15, 1996; accepted March 16, 1997}

% \abstract{}{}{}{}{} 
% 5 {} token are mandatory
 
\abstract
{We have measured the redshifts and single-aperture velocity dispersions of eight lens galaxies using the data collected by the Echellette Spectrograph and Imager (ESI) and Low Resolution Imaging Spectrometer (LRIS) at W.M. Keck observatory on different observing nights spread over three years (2018-2020). These results, combined with other ancillary data, such as high-resolution images of the lens systems, and time delays, are necessary to increase the sample size of the quasar-galaxy lens systems for which the Hubble constant can be measured, using the time-delay strong lensing method, hence increasing the precision of its inference. Typically, the 2D spectra of the quasar-galaxy lens systems get spatially blended due to seeing by ground-based observations. As a result, the extracted lensing galaxy (deflector) spectra become significantly contaminated by quasar light, which affects the ability to extract meaningful information about the deflector. To account for spatial blending and extract less contaminated and higher signal-to-noise ratio ($S/N$) 1D spectra of the deflectors, a forward modeling method has been implemented. From the extracted spectra, we have measured redshifts using prominent absorption lines and single aperture velocity dispersions using the penalized pixel fitting code \textsc{pPXF}. In this paper, we report the redshifts and single aperture velocity dispersions of eight lens galaxies - J0147+4630, B0445+123, B0631+519, J0659+1629, J0818-2613, J0924+0219, J1433+6007, and J1817+2729. Among these systems, six do not have previously measured velocity dispersions; for the other two, our measurements are consistent with previously reported values. Additionally, we have measured the previously unknown redshifts of the deflectors in J0818-2613 and J1817+2729 to be $0.866 \pm 0.002$ and $0.408 \pm 0.002$, respectively.}

\keywords{redshift -- velocity dispersion -- forward modeling -- time-delay            strong lensing method -- Hubble constant
               }
\authorrunning{P. Mozumdar et al.}
\maketitle
%-------------------------------------------------------------------
\section{Introduction}
The Hubble constant, $H_0$, denotes the current expansion rate of our Universe. \newadd{The value of $H_0$, though first measured almost a century ago, is still a matter of debate.} Utilizing cosmic microwave background (CMB) radiation, which is a relic from the early universe, the Planck collaboration has measured a Hubble parameter at the last scattering surface (z $\sim$ 1100) and then inferred a $H_0$ value of 67.4 $\pm$ 0.5 km s$^{-1}$ Mpc$^{-1}$ \citep{Planck_collab_2018} under the assumption of the standard cosmological model (cold dark matter with a cosmological constant, $\Lambda$CDM). However, using data from the local (or late) Universe and a cosmic distance ladder approach, where absolute magnitudes of type Ia supernovae in the Hubble flow have been calibrated by  Cepheids and parallax distances, the SH0ES team has measured $H_0 =$ 73.04 $\pm$ 1.04 km s$^{-1}$ Mpc$^{-1}$ \citep{Riess_2021b}. Several other independent probes also bolster this discrepancy between the model-dependent extrapolation and direct measurement of $H_0$ \citep[e.g.,][]{Abbott_2018, Aiola_2020, Freedman_2020, Pesce_2020, Kourchi_2020, Blakeslee_2021}. This unambiguous discrepancy is currently known as the Hubble tension \citep{Verde_2019, Abdalla_2022}. If systematic uncertainties in the measurement processes cannot account for this discrepancy, modifying the standard cosmological model or introducing new physics would be necessary. In this picture, more independent probes are crucial to resolving this issue. \\

One independent way to measure \ho\ is to use the time-delay strong lensing method \citep{Refsdal_1964}. When a time variable source such as a quasar or a supernova is strongly lensed by a foreground galaxy (deflector), multiple images of the background source form and the lensed light from these images arrive at different times. These relative delays depend on the gravitational potential (mass distribution) of the deflector, the large-scale mass distribution along the line of sight, and a combination of angular diameter distances involving the source, deflector, and observer\newadd{,} called the time-delay distance ($D_{\Delta t}$). By measuring high-precision time delays from multiple images and redshifts of the deflector and the source from spectroscopy, accurately modeling the mass distribution within the deflector, and estimating the extra-lensing deflection due to the large-scale mass distribution between the source and observer along the line of sight, one can infer $D_{\Delta t}$ \newadd{and angular diameter distance to the deflector ($D_d$),} and, hence constrain $H_0$ as it is inversely proportional to the time-delay distance \citep[e.g.,][]{Wong_2020, Millon_2020_TDCOSMO_I, TDCOSMO_IV, Rusu_2020, Shajib_2020, Chen_2019}. \referee{Since the first robust measurement of the time delays in the J0957+561 system, and the associated determination of $H_0$ \citep{Kundic_1997, Oscoz_1997, Oscoz_1996}, considerable improvements in the data quality and analysis techniques in all the aspects of the time-delay strong lensing approach over the last two decades have transformed this method into an effective and reliable tool to measure $H_0$.} %Considerable improvements in the data quality and analysis techniques in all the associated sectors over the last two decades have transformed the time\newadd{-}delay strong lensing method into an effective and reliable tool to measure $H_0$.
However, the well-known mass-sheet transform \citep[MST;][]{Falco_1985, Schneider&Sluse_2013} can still introduce significant uncertainty in this method {\newadd{\citep[e.g,][]{TDCOSMO_IV}}}. The MST is a mathematical degeneracy \newadd{which} implies \newadd{that} different mass distributions in the deflector can predict the same set of \newadd{imaging} observables (image position, flux ratios, etc.). As \referee{these different mass models lead} to different time\newadd{-}delay distances, the precision of the measured value of \ho\ suffers considerably. \referee{Assuming a specific mass model such as the power-law mass profile can induce a potential systematic in the \ho\ because such an assumption artificially breaks the MST \citep[e.g.,][]{Xu_2016, Sonnenfeld_2018, Kochanek_2020, Kochanek_2021}.} \referee{However, this} mathematical degeneracy \newadd{is mitigated} by introducing an independent tracer of mass, such as the stellar velocity dispersion of the deflector \citep{Treu&Koopmans_2002b, Koopmans_2003, Suyu_2014}. Therefore, the stellar kinematics of the deflector is \newadd{one of the} crucial ingredients to accurately measure \ho\ using time\newadd{-}delay cosmography \referee{when the maximal degeneracy in \ho\ due to the MST of the mass model is allowed to be explored} \citep[][]{TDCOSMO_IV, TDCOSMO_V}. \newadd{One can also impose tighter constraints on other cosmological parameters from joint inference of $D_{\rm d}$ and $D_{\Delta t}$ by combining the time-delay measurements and lens models with the stellar velocity dispersions of the deflectors \citep[e.g.,][]{Jee_2015, Jee_2016, Simon_2016, Holicow_IX_Simon, Shajib_2018}}.\\

Besides improving the robustness of the analysis methods, and checking for unknown systematics \citep{Millon_2020_TDCOSMO_I, Gilman_2020_TDCOSMO_III, Chen_2021_TDCOSMO_VI, Vyvere_2021_TDCOSMO_VII, Shajib_2022_TDCOSMO_IX, Matt_2022_TDCOSMO_X}, another ongoing effort of the Time-Delay COSMOgraphy (TDCOSMO) collaboration is to increase the sample size of lensed quasar systems \newadd{with measured $D_{\Delta t}$}. By mitigating the uncertainties at the population level, this large sample of quasar-galaxy lens systems with their associated data\newadd{, especially stellar velocity dispersion of the deflectors,} can help to achieve a sub-percent measurement of \ho\ \citep{TDCOSMO_V}. This coordinated effort covers observations and analyses of important properties. For example, measuring time delays requires monitoring of the source images' light curves with adequate sampling \citep[e.g.,][]{Courbin_2018, Millon_2020_Time_delay_data}. High-resolution imaging from the \textit{Hubble Space Telescope} (\textit{HST}) or ground-based adaptive optics (AO) instruments is crucial for lens modeling \citep[e.g.,][]{Wong_2017, Shajib_2019_modeling, Chen_2019}. High $S/N$ spectroscopic observations provide redshifts of the source and the deflector along with stellar kinematics of the deflector \citep[e.g.,][]{Suyu_2017}. Environmental studies of the lens systems to quantify the extra-lensing distortion also depend on photometric and spectroscopic observations \citep[e.g.,][]{Rusu_2017, Dom_2019}. As a part of this effort, in this paper, we report newly measured redshifts and single aperture velocity dispersions of the deflectors from eight quasar-galaxy lens systems --- J0147+4630, B0445+123, B0631+519, J0659+1629, J0818-2613, J0924+0219, J1433+6007, and J1817+2729. \\

Due to limitations associated with the ground-based observations, the 2D spectra of the quasar-galaxy lens systems are typically spatially blended, and it is a challenging task to extract \newadd{clean} and high $S/N$ 1D spectra of the deflectors. To overcome this issue, we have implemented a forward modeling method to extract 1D spectra of the deflectors and, provided the $S/N$ of the extracted spectra are sufficiently high, we have measured single aperture velocity dispersions and the redshifts. This paper is structured in the following way. In section \ref{sec:targets}, a brief description of the target lens systems and related information is provided. Next, the observational setup, data reduction pipeline, and how the 1D spectra have been extracted using the forward modeling method are described in section \ref{sec:obs_data_reduction}. In section \ref{sec:analysis}, the measurement process of the redshifts and the single aperture velocity dispersions of the deflectors are presented along with the results. Finally, sections \ref{sec:discussion} and \ref{sec:conclusion} discuss the reported results and conclusions respectively. \newadd{A flat $\Lambda$CDM cosmology with $H_0 = 70.0$ \hunit and $\Omega_{\rm m} = 0.3$ is adopted when necessary.
}
%----------------------start of observing info table------------
\renewcommand{\arraystretch}{1.3}
\begin{table*}[t]
\caption{\label{table:obs_sum}Summary of spectroscopic observation of the lens systems for which redshifts and stellar velocity dispersions of the deflectors have been reported in this paper. Position angle (PA) of the slits was measured from North to East direction. \newadd{The seeing conditions (FWHM) were estimated using the data from the corresponding nights.}} 
\centering                          % used for centering table
\begin{tabularx}{1.01\textwidth}{c c c p{1.9cm} c r r r c c}        % centered columns (4 columns)
\hline\hline                 % inserts double horizontal lines
Lens system & RA & Dec & Observation & Instrument & Total & Seeing & Slit & Slit & Slit\\   
&  &  & date &  & exposure &  & PA & width & Length\\
& (J2000) & (J2000) & (UT) &  & (sec) & (arcsec) & (degree) & (arcsec) & (arcsec)\\
\hline                        % inserts single horizontal line
J0147+4630 & 01:47:10.20 & +46:30:40.50 & Dec 1, 2018 &  ESI & 7200 & 0.70 & 7 & 1.0 & 20\\
 B0445+123 & 04:48:23.25 & +12:27:51.10 & Dec 13, 2020 &  LRIS & 12000 & 1.06 & 67 & 1.5 & 175\\ 
 B0631+519 & 06:35:11.70 & +51:56:48.50  & Nov 21, 2020 &  LRIS & 6080 & 0.69 & 315 & 1.0 & 175\\
 J0659+1629 & 06:59:03.82 & +16:29:08.90 & Mar 3, 2019 &  LRIS & 6800 & 1.33 & 65 & 1.0 & 175\\
 J0818-2613 & 08:18:28.24 & $-$26:13:24.80 & Apr 10, 2019 &  ESI & 3600 & 0.71 & 90 & 1.0 & 20\\
 J0924+0219 & 09:24:55.82 & +02:19:24.80 & Dec 1, 2018 &  ESI & 7200 & 0.70 & 10 & 1.0 & 20\\ 
 J1433+6007 & 14:33:22.80 & +60:07:15.60 & Apr 10, 2019 &  ESI & 7200 & 0.71 & 88 & 1.0 & 20\\
 J1817+2729 & 18:17:30.68 & +27:29:43.50 & Apr 10, 2019 &  ESI & 4800 & 0.71 & 315 & 1.0 & 20\\
\hline                                   %inserts single line
\end{tabularx}
\end{table*}
%------------------end of observing info table-------------------
\section{Targets}
\newadd{This section provides a brief description of the eight quasar-galaxy lens systems for which the deflector redshifts and velocity dispersions are measured.} The images of these lens systems along with the slit positions are presented in \newadd{Figure \ref{fig:image_targets}}.
\label{sec:targets}
\subsection{J0147+4630} 
This quadruple lensed quasar system was coincidentally discovered \citep{J0147_discovery_2017} in the Panoramic Survey Telescope and Rapid Response System \citep[Pan-STARRS1;][]{Pan-STARRS1_survey} survey data. The spectroscopic redshift of the quasar is $2.341 \pm 0.001$ \citep{J0147_quasar_z_Lee_2017}. The  redshift of the deflector is also measured, \zd \ $ = 0.678 \pm 0.001$, along with the velocity dispersion,  $313 \pm 14 $ \vdunit , by \citet{J0147_lens_z_veldis_Goicoechea}. \newadd{In this paper, we provide an independent measurement of the velocity dispersion of the deflector using a larger wavelength range than the one used in the previous measurement.} Lens modeling has been carried out by \citet{Shajib_2019_modeling} using  \textit{HST} imaging of the system.  
%and time delay between image A and B is also available at \citep{J0147_lens_z_veldis_Goicoechea}. 

\subsection{B0445+123}
This radio-loud double-imaged AGN was discovered by \citet{B0445_discovery_paper} in the Cosmic Lens All-Sky Survey \citep[CLASS;][]{CLASS_survey,CLASS_survey_2}. Radio observations have measured the separation between the lensed images to be 1.32\as .
%and \textit{HST} imaging is also available (proposal ID 9744, cycle 12).
Although  the AGN redshift is yet to be measured, spectroscopic observations have found that the deflector redshift is  \zd\ $= 0.5583\pm 0.0003$ \citep{McKean_keck_spec_2004}. 

\subsection{B0631+519}
This is another radio-loud double-imaged lens system discovered by CLASS \citep{B0631_discover_paper}. Observations at different radio wavelengths have revealed two compact lensed images separated by around 1.16\as. \textit{HST} imaging of the system has shown that there are two galaxies along the line of sight of the quasar. The main lensing galaxy is at redshift, \zd\ $ = 0.6196 \pm 0.0004$ and the other one is at redshift $0.0896 \pm 0.0001$ \citep{McKean_keck_spec_2004}.  However, the quasar redshift is yet unknown. 

\subsection{J0659+1629}
This quadruply-lensed quasar was first identified as a lens candidate by \citet{Delchambre_2019} using \textit{Gaia} data release 2 \citep{Gaia_DR2}. This detection was confirmed by \citet{Stern_2021} with measured source redshift, \zs \ $=3.083$ and deflector redshift, \zd \ $=0.766$. \newadd{The system was also independently discovered by \citet{Lemon_2022}. Lens modeling of the system is conducted by \citet{Schmidt_2022} using \textit{HST} imaging.}

\subsection{J0818-2613}
\newadd{\citet{Stern_2021} first discovered this quadruply imaged quasar system using \textit{Gaia} data release 2 and measured the quasar redshift, \zs \ $= 2.164$, using Keck-LRIS spectroscopy. This lens system was also independently reported by \citet{Lemon_2022} with a measured quasar redshift, \zs \ $= 2.155$. \citet{Schmidt_2022} performed lens modeling using \textit{HST} images of the system.} In this paper, we report the redshift and velocity dispersion of the deflector.
%To measure time delay currently Euler monitoring is ongoing (private communication). \textit{HST} imaging is available for this system (cycle 26)\\ 

\subsection{J0924+0219}
This quadruply lensed quasar was discovered using Sloan Digital Sky Survey \citep[SDSS;][]{SDSS_survey} data. \citet{Inada_2003_J0924_discovery} first reported this quad along with the measured source redshift \zs \ $ =1.524$, while \citet{Ofek_2006_J0924_zd} and  \citet{Eigenbrod_2006_J0924} measured the lens redshift, \zd \ $ =0.393$. Time-delay data between the two brightest images is available \citep{Millon_2020_Time_delay_data}. 
Lens modeling is presented by \citet{Eigenbrod_2006_J0924} using \textit{HST} imaging and recently by \citet{Geoff_2021_J0924_AO_modeling} using AO-based imaging.

\subsection{J1433+6007}
This quad was detected in the SDSS data release 12 photometric catalog \citep{SDSS_DR_11&12}. %\textit{HST} imaging shows that this system has a big Einstein radius \citep{Shajib_2019_modeling}.
\newadd{The source and deflector redshift, and deflector velocity dispersion \referee{were} first measured by} \citet{Agnello_2018_J1433}. %, and performed lens modeling.
The reported redshifts of the source and the deflector are, \zs \ $= 2.737\pm 0.003$ and \zd \ $= 0.407 \pm 0.002$, respectively. The measured velocity dispersion is $216 \pm 55$ \vdunit\ \newadd{using a 1\as\ aperture} \citep{Agnello_2018_J1433}. \newadd{However, this measurement only used the Ca K line and the spectrum was significantly contaminated by quasar light. In this paper, we present a new velocity dispersion measurement on a larger wavelength range and higher $S/N$ deflector spectrum.} \citet{Shajib_2019_modeling}  \newadd{and \citet{Schmidt_2022} both } performed lens modeling using \textit{HST} imaging.
%Besides, monitoring also going on to measure time delays between quasar images. 

\subsection{J1817+2729}
This quad was first identified as a lens candidate by \citet{Delchambre_2019} using \textit{Gaia} data release 2 and \newadd{then} independently confirmed with measured source redshift, \zs \ $=3.07$, by \citet{Lemon_2019, Stern_2021}. Detailed mass modeling has been conducted by \citet{Rusu_2018_J1817_modeling} using SUBARU/FOCAS imaging \newadd{and by \citet{Schmidt_2022} using \textit{HST} imaging}. In this paper, we present the deflector's redshift and velocity dispersion.

%Besides \textit{HST} image is also available (cycle 26). This system has an Einstein radius of 0.9 arcsec.

\begin{figure*}
    \centering
    {\includegraphics[width=\textwidth]{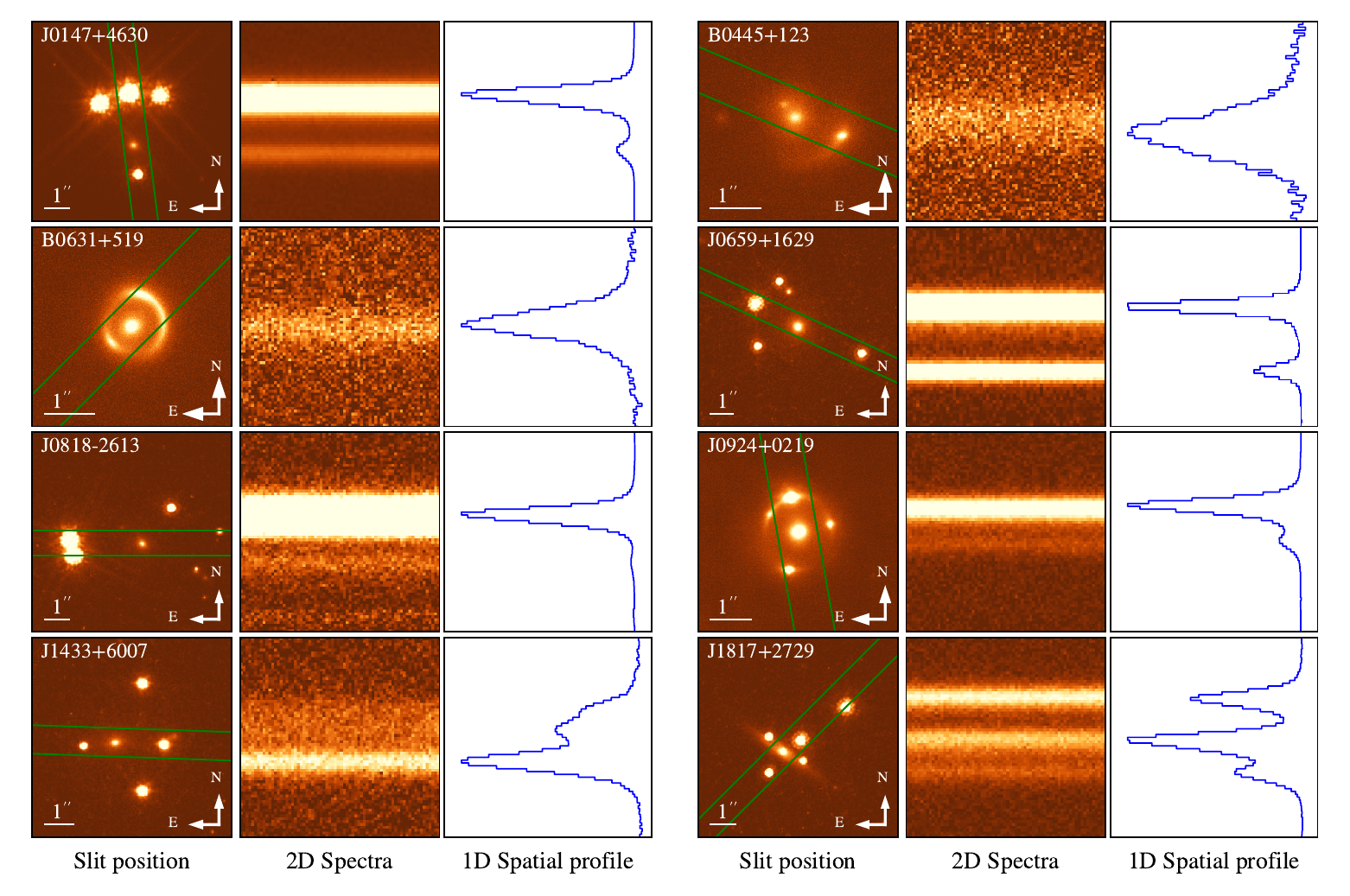}}
    \caption{Images of the lens systems, part of the collected 2D spectra, and the corresponding 1D spatial profiles. The 1D spatial profiles have been generated by collapsing the data in the 2D spectra along the spectral axis. The slit position (green) for each lens system is also shown. For the lens system, J0147+4630, J0659+1629, J0818-2613,  J1433+6007, and J1817+2729, we have used archival \textit{HST} images (PI: T. Treu) and for B04453+123, B0631+519 and J0924+0219, we have used ground based AO images (PI: C. D. Fassnacht).}
    \label{fig:image_targets}
\end{figure*}

%---------------start of figure including model--------------------
\begin{figure*}
    \centering
   {\includegraphics[width=\textwidth]{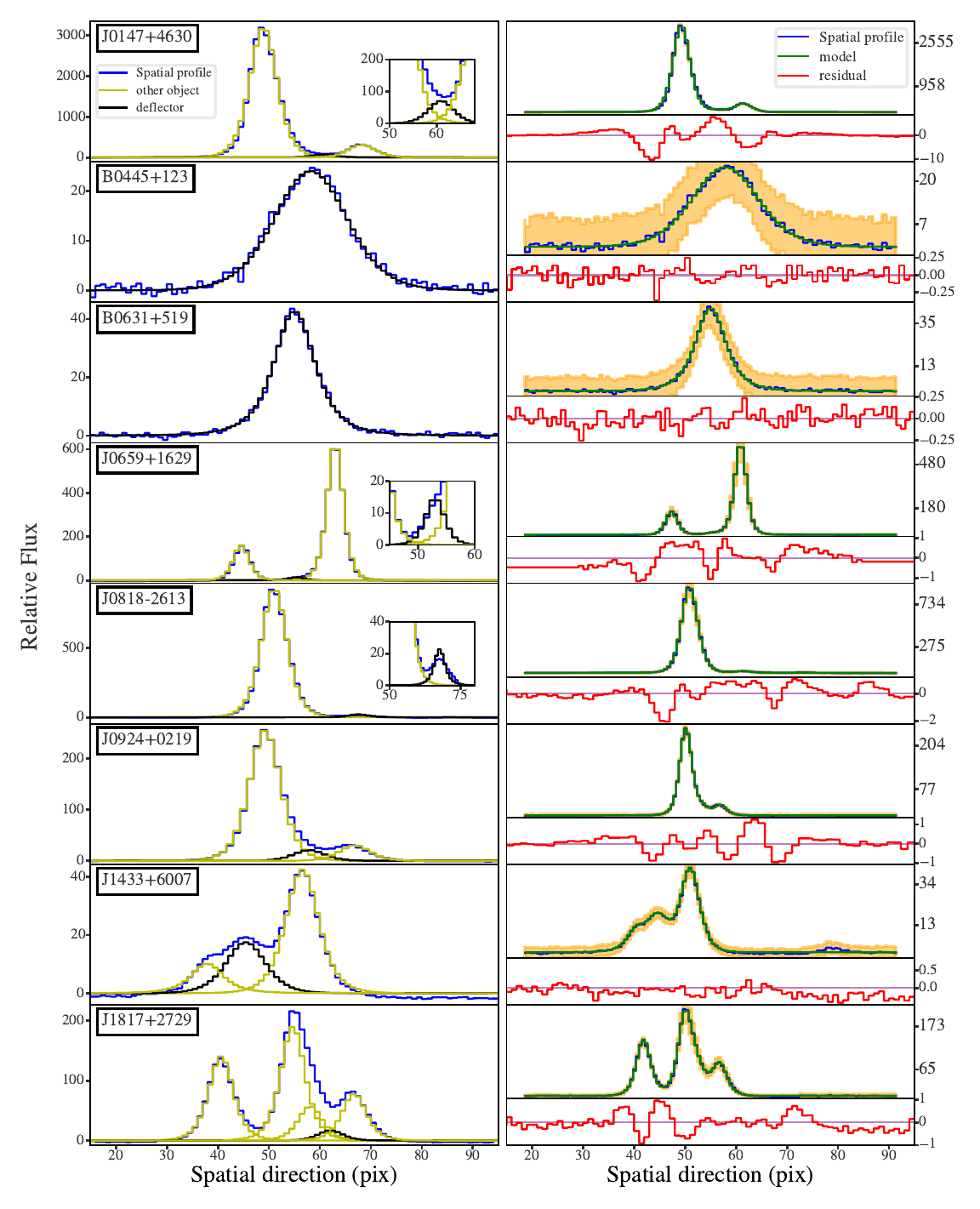}}
   \vspace*{-7mm}
      \caption{Illustration of model fitting to 1D spatial profiles. Left column - each row shows the 1D spatial profile (blue) with individual model components for a single lens system. The model components are Moffat profiles, \newadd{which represent the deflector and lensed quasars, with an additive polynomial function representing the background.} The profile component representing the deflector is marked black and other profile components generally representing the lensed quasar images are marked with yellow. Right column - The top panel in each row shows the 1D spatial profile (blue) with \newadd{1$\sigma$ uncertainty (orange) and} the corresponding model (green) that fits the data in that profile. The bottom panel shows the residual (red) between the data and the model \newadd{normalised to the uncertainties in data. \newadd{The uncertainties on the spatial profiles of B0445+123 and B0631+519 are higher compared to others because these 1D profiles were generated from single frames, while for other lens systems, the corresponding profiles were generated from coadded frames.}}}
     \label{Fig:modelfit_2}  
\end{figure*}

%---------------figure for the extracted spectra---------------------
\begin{figure*}
    \centering
    {\includegraphics[width=\textwidth]{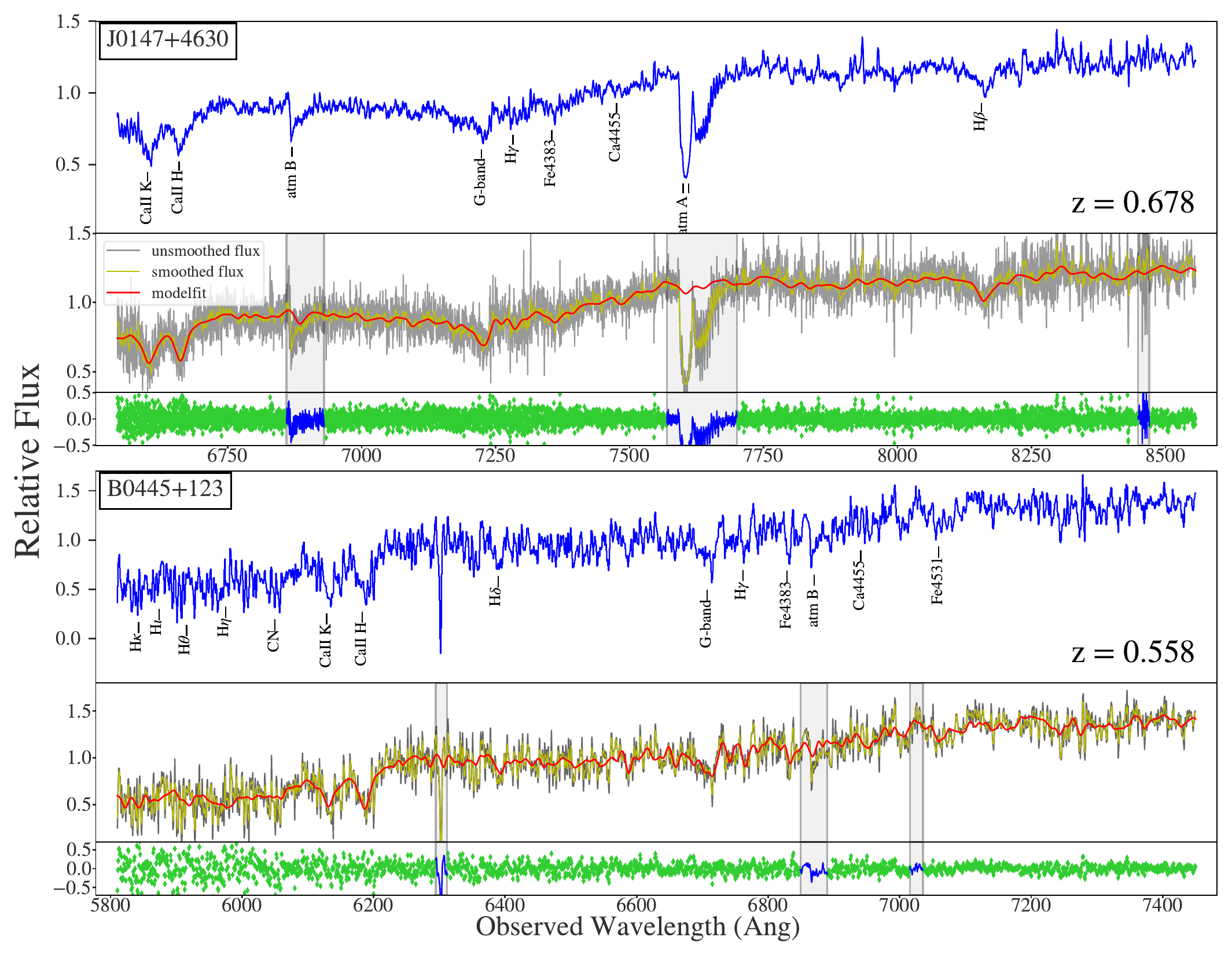}}
    \caption{1D spectra and velocity dispersion fits for the deflectors from the lens system J0147+4630 and B0445+123. The plot for each system consists of three panels. Top panel - the extracted and smoothed 1D spectrum of the deflector is plotted over the observed wavelength where the smoothing has been done using a \newadd{boxcar filter} of size around 2.5 \r{A}. Prominent stellar and telluric absorption lines are marked if present. And the measured redshift of the deflector is mentioned in the lower right corner of the plot. Middle panel - the \textsc{pPXF} generated model fit (red) to measure the velocity dispersion is plotted on top of the original unsmoothed spectrum (black). A smoothed version (yellow) of the original spectrum is also presented to show the goodness of the fit. The grey parts are the masked or excluded regions from the fit. Bottom panel - the green dots show the residuals between the unsmoothed flux and the model fit for each wavelength \newadd{normalized to the respective model fits.}}
    \label{fig:extracted_spec1}
\end{figure*}

\begin{figure*}
    \centering
    {\includegraphics[width=\textwidth]{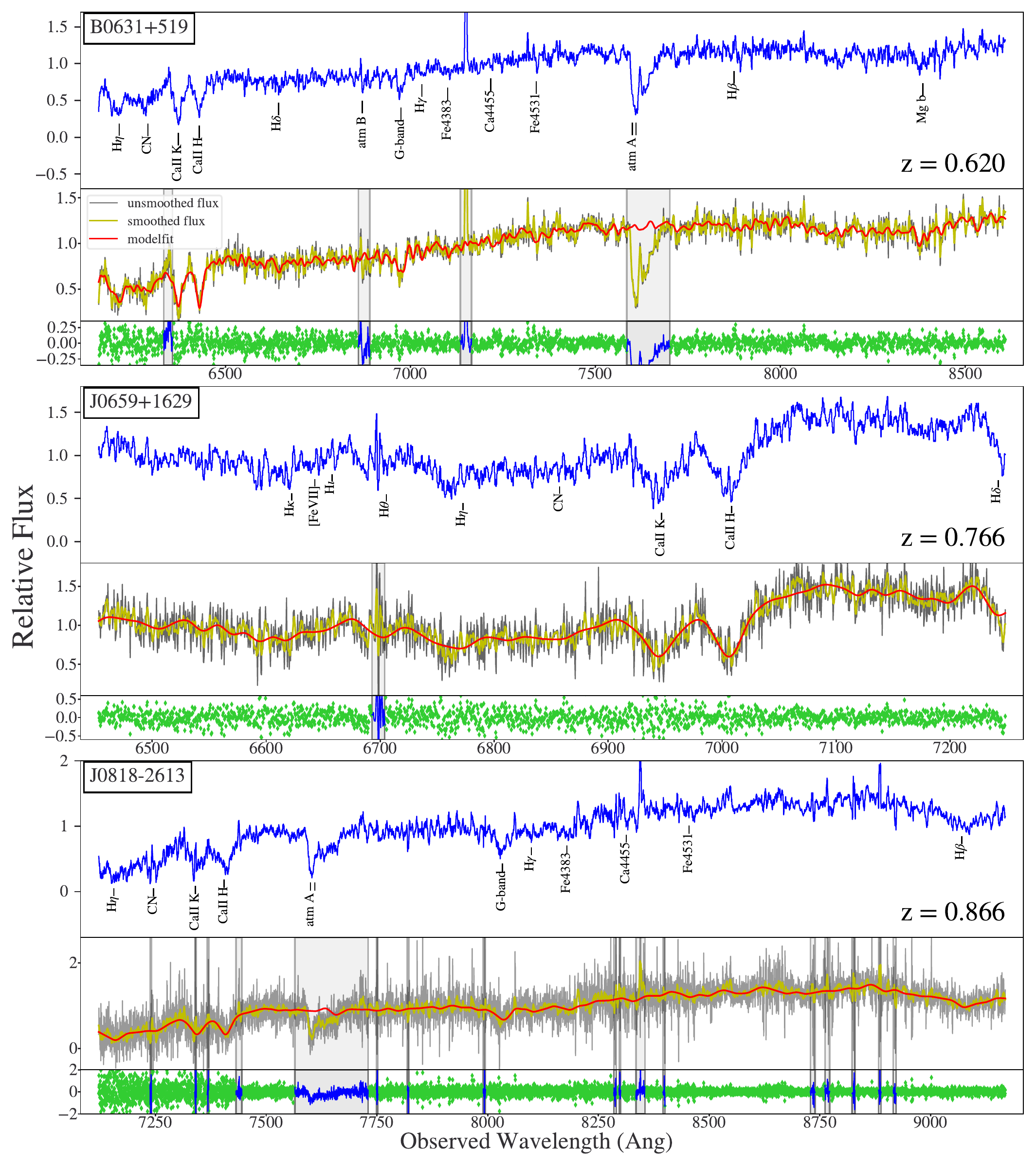}}
    \caption{ 1D spectra and velocity dispersion fits for the deflectors from the lens system B0631+519, J0659+1629 and J0818-2613. The description of the figure is same as Figure \ref{fig:extracted_spec1}.}
    \label{fig:extracted_spec2}
\end{figure*}

\begin{figure*}
    \centering
    {\includegraphics[width=\textwidth]{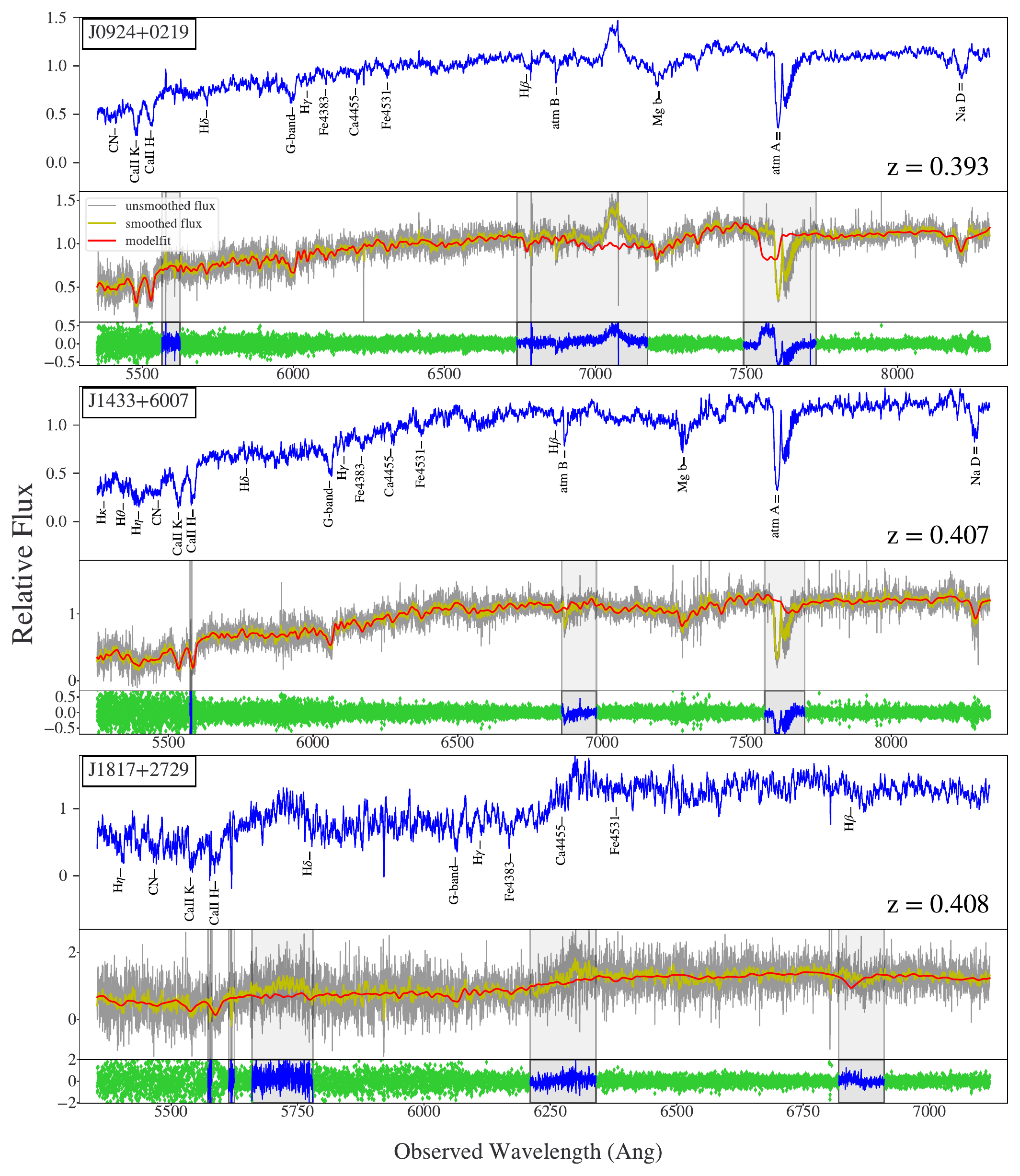}}
    \caption{ 1D spectra and velocity dispersion fits for the deflectors from the lens system J0924+0219, J1433+6007 and J1817+2729. The description of the figure is same as Figure \ref{fig:extracted_spec1}.}
    \label{fig:extracted_spec3}
\end{figure*}

\section{Observations and data reduction}
\newadd{In this section, we first describe the observations for our sample of eight lens systems, which include two different spectrographs and three different setups. Next, the data reduction pipelines are described. Finally, we present the forward modeling method and how the 1D spectra were extracted from the respective 2D spectra. }
\label{sec:obs_data_reduction}
\subsection{Observations}
The spectroscopic observations for the target lens systems were carried out using the Low Resolution Imaging Spectrometer (LRIS; \citealt{LRIS_Oke_1995}) and the Echellette Spectrograph, and Imager (ESI; \citealt{ESI_Sheinis_2002}) mounted on the 10m Keck telescopes. The observing nights were spread over three years, from 2018 to 2020, with typical seeing conditions (FWHM) ranging $\sim$ 0.7 \as\ - 1.33 \as. 
The ESI spectra were taken in Echellette mode using the 1.0\as\ wide slit \referee{which provides a spectral resolution $R \approx 4000$}. %, that translates in velocity space to $\Delta v = c/R \simeq 74.75$\vdunit). 
This setup covers a wavelength range from 3900 Å to 10900 Å with a constant dispersion 
%of $1.65 \times 10^{-5}$ \r{A}/pixel 
in \newadd{logarithmic} space corresponding to 11.5 \vdunit\ in velocity, while the pixel scale in the spatial direction varies from 0.120\as\ in the bluest order to 0.168\as\ in the reddest order. The LRIS spectra were collected in the long-slit mode using both blue and red arms simultaneously.
%and thus achieving a total wavelength coverage of 3050-9,000 \r{A}. 
A dichroic was used to split the light beam at 5696 \r{A}. The blue side was configured with a 600/4000 grism that provides a 0.63 \r{A}/pixel dispersion and covers a wavelength range of 3040-5630 \r{A} \referee{with a spectral resolution of $R \approx 1100$ at central wavelength $\lambda \approx 4340 \r{A}$}. On the red side, two different configurations were used. To observe the lens systems B0445+123 and B0631+519, a 600/7500 grating was used, which produces a dispersion of 0.8 \r{A}/pixel while covering a wavelength range of 5600-9000 \r{A}. For the lens system J0659+1629, the 1200/7500 grating was used, which covers a wavelength range of 5600-7250 \r{A} with a dispersion of 0.4 \r{A}/pixel. \referee{Both setups on the red side provide a spectral resolution of  $R \approx 1400$}. The plate scales on the blue and the red side are 0.123\as/pixel and 0.135\as/pixel, respectively. The position angle (PA) of a slit was generally chosen in such a way that the brightest quasar image of the corresponding lens system fell within the slit along with the deflector and at least another quasar image. The total integration time for each system was divided typically into $n \times 2400$ \textrm{s} or $n \times 1800$ \textrm{s} or $n \times 1200$ \textrm{s} exposures. A summary of the main observing information is provided in Table \ref{table:obs_sum}.

\subsection{Data reduction}
The LRIS data for the lens system B0445+123 and B0631+519 were reduced using the \textsc{PypeIt} \citep{PypeIt_2020} pipeline and for J0659+1629 using the \textsc{LPipe} \citep{lpipe_2019} pipeline. These packages performed overscan subtraction, bias and dark current correction, and flat-fielding. Finally, the packages produced sky-subtracted 2D science spectra and the corresponding variance and wavelength data. \textsc{PypeIt} provides 2D wavelength images while \textsc{LPipe} generates 1D wavelength solutions. \referee{The pixels contaminated with cosmic rays were marked by comparing each frame to the respective median frame. A median frame for each system was created by combing all the 2D image frames from that system.} After that, a rectification was applied to all reduced 2D data to account for any possible tilts present in the spectra, using a spline interpolation method. These rectified science, variance, and wavelength data were used to extract the 1D spectra\newadd{, as described in the next section.} \\

The ESI data were processed via a custom python-based data reduction package.  The code automatically locates the ten individual spectral orders, performs a bias subtraction and flat field correction on the data, rectifies the orders, subtracts the sky emission, and performs a wavelength calibration.  \referee{The cosmic-ray rejection for the ESI data was performed by running edge-detection algorithms that our tests have shown to be effective for this process.} The output of the code includes the calibrated and background-subtracted data for each exposure and the corresponding variance spectra. \\

For B0445+123 and B0631+519, the 1D deflector spectra were extracted from each 2D science frame and then coadded using an inverse-variance weighting process to get the final 1D spectra. \newadd{As the 2D wavelength solutions for these two systems varied from frame to frame, it was easier to handle the issue when the 1D spectra were extracted from each frame.} For the remainder of the lens systems, first, the individual 2D spectra were coadded with a inverse variance weighting, and then the 1D deflector spectra were extracted from the coadded 2D spectra.

%-------------measured redshifts------------
\renewcommand{\arraystretch}{1.3}
\begin{table*}[th]
\caption{Measured redshifts of the deflectors with associated uncertainties using the extracted 1D spectra. Previously known values, if any also given with corresponding references.}             % title of Table
\label{table:redshift}      % is used to refer this table in the text
\centering                          % used for centering table
\begin{tabular}{p{2cm} l c c}        % centered columns (4 columns)
\hline\hline                 % inserts double horizontal lines
Lens system & Redshift ($z_d$) & Previous measurement & References\\   
\hline
  J0147+4630 & 0.678  $\pm$ 0.0004 & 0.678$\pm$ 0.001 & \cite{J0147_lens_z_veldis_Goicoechea} \\
  B0445+123 & 0.558  $\pm$ 0.001 & 0.5583$\pm$ 0.0003 & \cite{McKean_keck_spec_2004} \\
  B0631+519 & 0.620  $\pm$ 0.001 & 0.6196$\pm$ 0.0004 & \cite{McKean_keck_spec_2004} \\
  J0659+1629 & 0.766 $\pm$ 0.0015 & 0.766 & \cite{Stern_2021} \\
  J0818-2613 & 0.866  $\pm$ 0.002 & None & \\      % inserting body of the table
  J0924+0219 & 0.393  $\pm$ 0.0008 & 0.393 & \cite{Eigenbrod_2006_J0924} \\ 
  J1433+6007 & 0.407  $\pm$ 0.0006 & 0.407$\pm$ 0.002 & \cite{Agnello_2018_J1433} \\
  J1817+2729 & 0.408  $\pm$ 0.002 & None & \\
\hline      %inserts single line
\end{tabular}
\end{table*}
%------elocity dispersion results table------
%\renewcommand{\arraystretch}{1.2}

\subsection{Forward modeling and 1D extraction}
The angular separations between the quasar images and the deflector in a lens system are typically on the order of 1\as. Hence, in seeing limited ground-based observations, the 2D spectra of the lens systems become spatially blended. This situation gets more complicated as the deflectors are generally orders of magnitude fainter than the bright quasar images. For example, in Figure \ref{fig:image_targets} the 2D spectra of J0147+4630, J0924+0219, and J1817+2729 do not show any distinct trace from the deflector. In this situation, it is challenging to extract uncontaminated and high $S/N$ 1D spectra of the deflectors. Traditional optimal extraction codes based on the algorithm of \citet{Horne_1986} are not up to the task as they \newadd{do not} account for the blending or cross-contamination. To overcome this challenge, we have implemented a forward modeling method \citep[e.g.,][]{Sluse_2007, Shalyapin_2017, J0147_lens_z_veldis_Goicoechea}. First, a \newadd{global} model is created that emulates the data in a 1D spatial profile of the 2D science spectrum. The 1D spatial profile is constructed by collapsing part of the 2D spectrum, typically containing a spectral width of 300-400 pixels, along the spectral axis. The model consists of several Moffat \citep{Moffat_1969} profiles \newadd{along with an additive polynomial function}. Each \newadd{component} profile \newadd{in the model} represents either a quasar image or the deflector that is \newadd{wholly} or partially captured by the slit and therefore has a significant contribution to the total light distribution across the slit, \newadd{while} the polynomial models the background. \newadd{The model} is further refined by constraining the parameters of the profiles in the following way. The projected center to center distances of the objects along the slit are fixed based on \newadd{the corresponding measured distances between the objects using} archival \textit{HST} images or AO-based images of the systems. We assume that the same point spread function (PSF) is applicable for all objects in the slit, and therefore they require that the profiles have the same shape parameters. \newadd{This is a reasonable assumption for the galaxies given their point-like sizes in the images of the lens systems, especially when considering the effects of seeing.} Figure \ref{Fig:modelfit_2} shows a refined global model for each lens system along with individual profile components. This figure also shows \newadd{the residuals normalized to the uncertainties in the data of the corresponding 1D spatial profiles.} Once the initial global description of the spatial profile has been determined, we take a multi-step approach to determine the component fluxes at each wavelength. \newadd{The spatial profile in a single column of the detector, that is, at a specific wavelength, is typically too noisy to do a full model fit to that single profile. Therefore, we take the following steps to use binned data to determine the model parameter values in each individual column. (1) We divide the 2D spectrum into spectral bins of equal widths, typically around 20-25 spectral pixels, and construct a series of 1D spatial profiles by summing the 2D data within these spectral-bins. Then we fit the global model to these spatial profiles. From the fitted models, we get a set of values for each centroid and shape parameter in the global model.} (2) \newadd{We fit a low-order polynomial as a function of wavelength to each set of fitted centroid and shape parameters values to describe their wavelength-dependence.} (3) \newadd{We create an individual model similar to the global model for each wavelength using the above-mentioned fitted polynomials, where all the parameters except the amplitudes are kept fixed, and then we fit the amplitudes.
%Then we fit the data in each spectral pixel or wavelength using the respective model for that wavelength.
%Then a variance weighted fitting is performed to the data in each spectral bin of one pixel using the respective model for that pixel. Finally, 
(4) We obtain the total flux for each object (quasar or galaxy) at each wavelength by integrating the corresponding fitted Moffat profile along the spatial direction. %We simultaneously extract 1D spectra for each component profile in the model by \newadd{integrating the flux in the profiles of individual components.} %converting the fitted amplitudes to flux.
We also calculate the associated variance on flux using the covariance matrix generated in the fitting process. }\\

All the extracted 1D spectra have been response-corrected using a spectrophotometric standard stellar spectrum collected by the same spectrograph on the corresponding observing night and extracted following the procedure described above. In Figures \ref{fig:extracted_spec1} to \ref{fig:extracted_spec3}, the extracted and smoothed deflector spectra are presented with the absorption and telluric lines marked. \newadd{The smoothing has been done using a boxcar filter of size around 2.5 \r{A}.}

\section{Data analysis}
\newadd{In this section, we discuss the redshift and velocity dispersion measurements of the deflectors and their associated uncertainties. We also discuss how the $S/N$ of the spectra were measured and describe a test to check the covariance of the velocity dispersion measurements. }
\label{sec:analysis}
\subsection{Redshift measurement}
Among the eight lensing galaxies presented in this paper, only the redshifts of the deflectors in J0818-2613 and J1817+2729 were not previously measured. The aforementioned forward modeling method enabled us to extract the deflector spectra for both of these systems with sufficient $S/N$ where several absorption lines such as $\mathrm{C}$a II  $\mathrm{K}$ and  $\mathrm{H}$ and the $\mathrm{G}$-band are clearly visible after smoothing (Figures \ref{fig:extracted_spec2} and \ref{fig:extracted_spec3}). The redshifts are presented in Table \ref{table:redshift}. The reported uncertainties are the standard deviations of the redshifts measured using several absorption lines such as $\mathrm{CN}$, $\mathrm{C}$a II  $\mathrm{K}$ and  $\mathrm{H}$, $\mathrm{G}$ band, $\mathrm{H \beta}$, $\mathrm{H \delta}$, $\mathrm{Mg}$ b etc. Our measured redshifts for the remainder of the six deflectors are consistent with the previously known values.

\renewcommand{\arraystretch}{1.3}
\begin{table*}[t]
\caption{Measured $S/N$ per angstrom of the extracted deflector spectra and single aperture velocity dispersion with associated statistical and systematic uncertainties. As a sanity check, SIS velocity dispersions are presented to compare with the measured velocity dispersions. The quasar redshifts and Einstein radii reported here have been used to calculate the SIS velocity dispersions. Note that, for B0445+123 and B0631+519, the source redshift is unknown. Thus a possible range of source redshift has been assumed to calculate the corresponding SIS velocity dispersion range.}
\label{table:veldis_measurements}
\centering                          % used for centering table
\begin{tabular}{c c c c c c l c}        % centered columns (4 columns)\begin{table*}[ht!]
\hline\hline                 % inserts double horizontal lines
Lens system & $S/N$ & Velocity  & Statistical & Systematic & Source & Einstein & SIS velocity   \\ 
 & per \r{A} & dispersion & uncertainty & uncertainty & redshift ($z_s$) &  radius & dispersion  \\ 
 & & (\vdunit) & (\vdunit) & (\vdunit) &  & (arcsec) & (\vdunit) \\
\hline                        % inserts single horizontal line
 J0147+4630 & 12 & 283 & 30 & 12 & 2.377 & 1.886\tablefootmark{a} &  339 \\
 B0445+123 & 12 & 136 & 21 & 13 &  2.0 - 3.5& 0.675\tablefootmark{b} & 198 - 184 \\ 
 B0631+519 & 18 & 147  & 9 & 10 & 2.0 - 3.5 & 0.58\tablefootmark{b} & 190 - 174 \\
 J0659+1629 & 13 & 326 & 28 & 13 & 3.083 & 2.124\tablefootmark{a} &  356 \\
 J0818-2613 & 7 & 392 & 46 & 22 & 2.15 & 2.896\tablefootmark{a} & 472 \\
 J0924+0219 & 23 & 209 & 9 & 12 & 1.523 & 0.94\tablefootmark{c} & 223 \\ 
 J1433+6007 & 20 & 261 & 6 & 7 & 2.74 & 1.581\tablefootmark{a} & 272 \\
 J1817+2729 & 5 & 198 & 45 & 25 & 3.07 & 0.893\tablefootmark{a} & 203 \\
\hline                                   %inserts single line
\end{tabular}
\tablefoot{
\tablefoottext{a}{Mean value of Einstein radius obtained by lens modeling in \citet{Schmidt_2022} using \textit{HST} images.} \\
\hspace*{0.8cm} \tablefoottext{b}{As detailed lens modeling for this system isn't available, half of the quasar image separation measured from AO-based images was \hspace*{1.1cm}considered Einstein radius.} \\
\hspace*{0.8cm} \tablefoottext{c}{Mean value of Einstein radius obtained by lens modeling in \citet{Geoff_2021_J0924_AO_modeling} using AO-based images.}
}
\end{table*}
\renewcommand{\arraystretch}{1.3}
%---------------end of velocity dispersion result table---------------

\subsection{Velocity dispersion measurement}
\newadd{To achieve reliable velocity dispersion measurements, high $S/N$ spectra are necessary. The mean $S/N$ per angstrom for all the extracted deflector spectra except J0659+1629 were calculated using rest-frame wavelength ranges 4150 - 4250 \r{A} and 4630 - 4760 \r{A}. These two wavelength ranges were chosen based on the lack of any prominent absorption or telluric lines and also because none of the spectra have any emission lines or other contamination present within these ranges. However, as these rest-frame wavelength ranges were not available for the extracted deflector spectrum of J0659+1629 system, a separate rest-frame wavelength range from 3990 \r{A} to 4090 \r{A} was used to calculate the $S/N$ per angstrom. The estimated $S/N$ of the deflector spectra are reported in Table \ref{table:veldis_measurements}.}\\

To measure the line-of-sight (LOS) stellar velocity dispersions of the lensing galaxies, we have used a penalized pixel fitting method implemented through an in-house analysis pipeline \textsc{veldis} \footnote{\url{https://github.com/pmozumdar/veldis}} which is a wrapper around the \textsc{pPXF} package \citep{Cappellari_2004, Cappellari_2017_ppxf}. This method measures the velocity dispersion by fitting a model to the observed galaxy spectrum in pixel space. The models are created using a weighted linear combination of the broadened stellar templates to which a sum of orthogonal polynomials is added. The additive polynomials are used to adjust the continuum shape of the templates during the fit. 
%As we are only interested in measuring the Doppler broadening of the observed absorption lines, we first need to account for the instrumental broadening. This process is performed by applying a Gaussian smoothing to the stellar templates as the line spread function (LSF) of a spectrograph can be fairly described by a Gaussian distribution \citep{Cappellari_2017_ppxf}. The spread of the Gaussian kernel used for smoothing is proportional to the offset between the FWHMs of the LSF of the spectrograph used to observe the template spectra and the LSF of our science observation.
As templates we have used the Indo-US stellar library \citep{Indo_US_lib}, which consists of 1273 stellar spectra covering a wavelength range of 3460 - 9464 \r{A} with constant dispersion of 0.4 \r{A} and FWHM resolution $\sim$ 1.35 \r{A} \citep{MILES_library_2011}. The FWHM resolutions of the instruments used to collect our spectra have been measured by fitting Gaussian profiles to several emission lines at varying wavelengths from the corresponding sky spectra. \referee{Though the measured FWHM resolutions depend on wavelength, the range of resolutions is not significant enough to affect the velocity dispersion measurement. Thus we have adopted a constant FWHM resolution and ignored its wavelength dependence.} As prescribed in \citet{Cappellari_2017_ppxf}, we have shifted the galaxy wavelengths and FWHM resolutions of the galaxy spectrographs to the rest frame, as the redshifts of the observed deflectors are relatively high. \newadd{In the observed frame, the FWHM of all our observations was higher than that of the templates. However, Gaussian smoothing was carried out on the template spectra only if the observed FWHM was higher than the templates' after shifting to the rest frame.} 
%After that, if the FWHM resolutions of our observations are still higher than the templates', we carried out the Gaussian smoothing on the template spectra.
For all the galaxy spectra, we have masked the atmospheric A-band and B-band absorption lines, if present. Emission lines from the foreground galaxies and contaminated regions due to quasar light, if any, have also been excluded from the fit. Figures \ref{fig:extracted_spec1} to \ref{fig:extracted_spec3} show all the fits generated by \textsc{pPXF}. In these plots, the gray-colored spectra are the unsmoothed, logarithmically rebinned deflector spectra, the red-colored spectra are the best fit models, and the yellow-colored spectra are the smoothed versions of the original spectra. The vertical shaded regions mark the excluded part of the deflector spectra from the fitting. The green dots show the residuals between the unsmoothed flux and the model fit for each wavelength normalized to the respective model fits. \newadd{Unlike the optimal extraction-based methods, the extraction technique presented in this paper and others \citep[e.g.,][]{J0147_lens_z_veldis_Goicoechea, Melo_2021} doesn't require defining an aperture along the spatial direction. Thus, the measured velocity dispersions correspond to the integrated velocity dispersions within a rectangular spectroscopic aperture of the respective slit-widths and slit-lengths centered on the deflectors (see Table \ref{table:obs_sum}).} After obtaining the velocity dispersion measurements from the fits, we have estimated the statistical and systematic uncertainties associated with the corresponding measurements. To determine the statistical uncertainty, we have performed three hundred Monte-Carlo simulations by adding Gaussian noise to the \newadd{smoothed} deflector spectra. The noise at each wavelength has been generated using normal distributions with variances equal to the variances of the corresponding galaxy spectra \newadd{while ignoring the pixel-to-pixel covariances, as they are negligible compared to the respective variances.} The fits are carried out with the same setup used to measure the respective velocity dispersion reported here. We measured the systematic uncertainties by conducting roughly three hundred fits while varying the wavelength range of the observed galaxy spectrum that was used for the measurement, the subset of templates, and the degree of the additive polynomials. The measured velocity dispersions of the deflectors along with the associated statistical and systematic uncertainties are reported in Table \ref{table:veldis_measurements}. \\

\newadd{ While fitting for the velocity dispersion, we have attempted to include as many absorption lines as possible without compromising the respective $S/N$ of the deflector spectra. However, the $S/N$ levels of the extracted spectra affect the measured velocity dispersions significantly. From the reported uncertainties in Table \ref{table:veldis_measurements}, it is clear that both the statistical and systematic uncertainty increase noticeably as the $S/N$ levels decrease. On the contrary, both of these uncertainties improve when the $S/N$ levels of the extracted spectra are around 20 or above, as is the case for the deflectors in the B0631+519, J0924+0219, and J1433+6007 lens system. In any case, all the extracted spectra are mostly uncontaminated by the quasar light, and the absorption features are sufficiently strong enough to conduct a velocity dispersion measurement.} \\

\begin{figure*}[th!]
    \centering
    {\includegraphics[width=\textwidth]{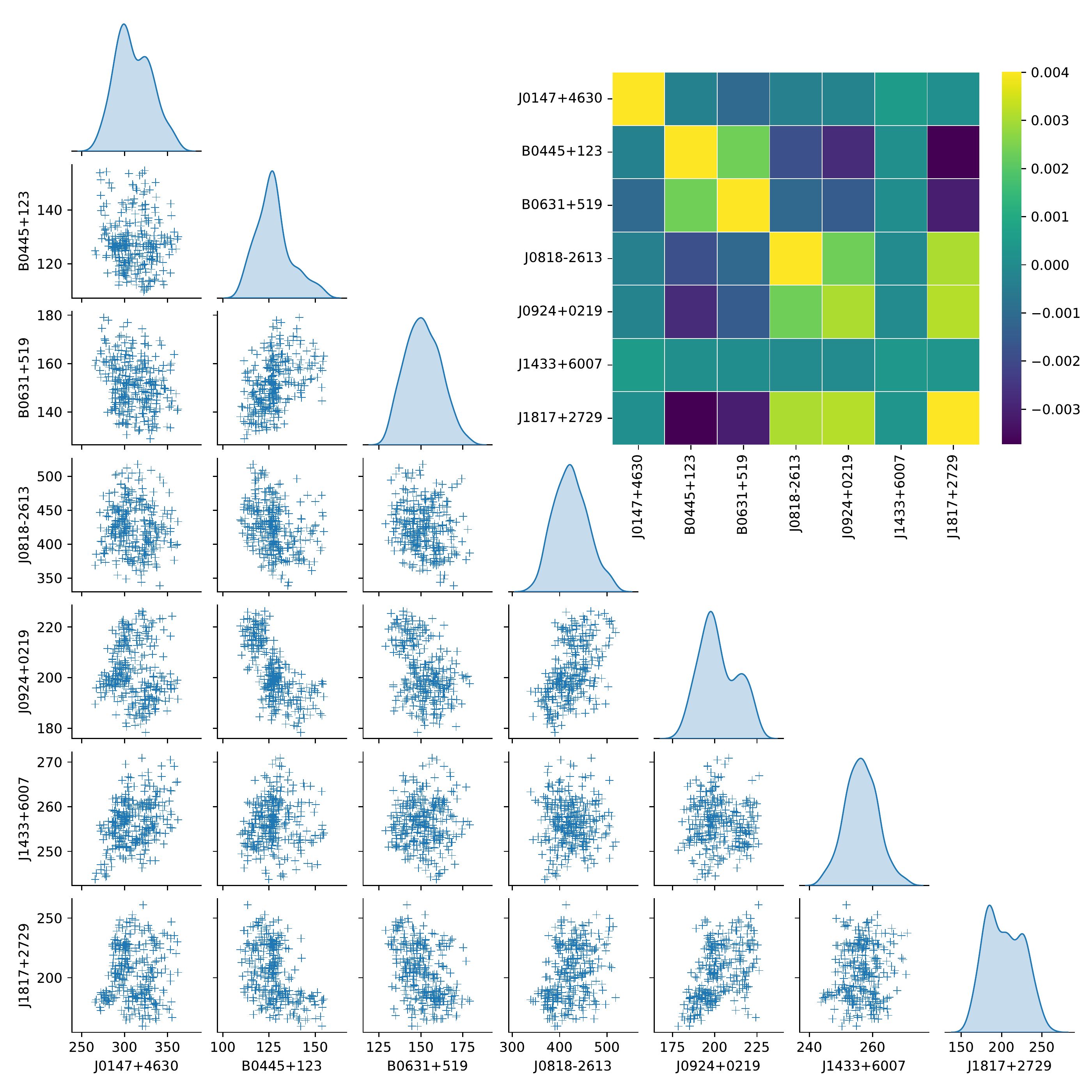}}
    \caption{ \newadd{Illustration showing relations among the velocity dispersion measurements of the deflectors. The scatter plots between each pair of deflectors show corresponding velocity dispersion measurements for three hundred setups where each marker shows the measurements from the same setup. \referee{The blue-shaded curves are the individual velocity dispersion distributions of the deflectors.} The set-ups were varied by changing the wavelength ranges, set of the spectra of the templates, and degree of the additive polynomial. In general, the covariance terms are negligible except in a few cases, for which they should be taken into account. The lens system J0659+1629 has been excluded from this analysis as the extracted deflector spectrum for this system only covers a small wavelength range. The plot in the upper-right shows a heat-map of the covariance matrix calculated from these velocity dispersion measurements, where the measurements were normalized to their respective mean.}}
    \label{fig:covariance_corner}
\end{figure*}

\newadd{We have also checked whether the velocity dispersion measurements of the deflectors vary preferentially in the same direction given the same set-up, e.g., the same rest-frame wavelength range, the same set of templates, etc. The lens system J0659+1629 has been excluded from this analysis as the extracted deflector spectrum for this system only covers a small wavelength range and thus provides almost no scope to vary the wavelength. With the rest of the deflectors, we have performed around three hundred sets of fits under different set-ups while using the same set-up for each group of seven fits corresponding to seven deflectors. The set-ups were varied by changing the wavelength ranges, the templates used, and the degree of the additive polynomial. These measurements and how they co-vary for each pair of deflectors are shown as a corner plot in Figure \ref{fig:covariance_corner}. The figure also presents the heat-map of the covariance matrix calculated from these velocity dispersion measurements, where the measurements were normalized to their respective mean. From this figure, one can see that except in a few cases, there is almost no correlation among the velocity dispersion measurements of the seven deflectors. However, in those cases the off-diagonal terms in the covariance matrix are essential and can not be ignored.}

\section{Discussion}
\label{sec:discussion}

\subsection{Comparison to previous redshift and velocity dispersion measurements}
The redshifts of the six deflectors, which were previously known, match our measurements \newadd{to within 1$\sigma$}. \newadd{Among the eight lens systems, the deflector velocity dispersions of the J0147+4630, and J1433+6007 lens systems were previously measured. For the deflector in J0147+4630, the previous measurement is $313 \pm 14$ \vdunit\ using a 0.5\as -width slit under subarcsecond seeing conditions \citep{J0147_lens_z_veldis_Goicoechea}. This result is within the uncertainty of our measurement of $283 \pm 30 \pm 12$ \vdunit\ \referee{(see Table \ref{table:veldis_measurements})}. However, using the same wavelength range (6000 - 7500 \r{A}) as in the previous measurement, but with the setting used in this paper, we obtained a velocity dispersion value of 311 \vdunit. Hence, the difference between the measurements in their means may arise from the difference in the used wavelength ranges, aperture sizes and seeing conditions, etc. For the deflector in J1433+6007, the previous measurement is $216 \pm 55$ \vdunit\ by fitting only to Ca K line \citep{Agnello_2018_J1433}. The data for this measurement was collected using a 1\as\ wide spectroscopic aperture under $\sim$ 1\as\ seeing. Our measured velocity dispersion for this deflector is $261 \pm 6 \pm 7$ \vdunit\ \referee{(see Table \ref{table:veldis_measurements})}, which is higher but still within the uncertainty of the previous measurement.}

\subsection{Comparison to predictions from SIS lens models}
The strength of the absorption lines, especially $\mathrm{C}$a II  $\mathrm{K}$ and $\mathrm{H}$, presence of 4000 \r{A} break and the lack of emission lines in the extracted spectra confirm that the deflectors of the lens systems reported in this paper are early-type galaxies. \newadd{Thus as a sanity check, we have compared measured velocity dispersion values with theoretically derived ones by assuming a singular isothermal sphere (SIS) profile for the galaxies.}
%This information provides us with an opportunity to compare our measured velocity dispersion values with. For the singular isothermal sphere (SIS) profile which
%This mass profile is the simplest mass distribution of an extended system, \newadd{and is roughly consistent with typical mass distributions observed in nearby ellipticals.
\newadd{This mass profile is the simplest choice to predict the velocity dispersion using only a single lens model parameter, that is the Einstein radius. Under this assumption,} the single-aperture velocity dispersion of the system $\sigma_{\rm SIS}$ \newadd{can be calculated} as -
\begin{equation*}
  \sigma_{\rm SIS} = \sqrt{\frac{c^2}{4\pi}\theta_{\rm E} \frac{D_{\rm s}}{D_{\rm ds}}}
\end{equation*}
where $\theta_{\rm E}$ is the Einstein radius in \newadd{arcseconds}, $D_{\rm s}$ and $D_{\rm ds}$ are the angular diameter distances between the observer and the source, and the deflector and the source, respectively, and $c$ is the speed of light. \newadd{For lens systems J0147+4630, J0659+1629, J0818-2613, J1433+6007, and J1817+2729, the mean value of Einstein radius obtained by lens modeling in \citet{Schmidt_2022} was used, and for the system J0924+0219, the mean value of Einstein radius obtained by lens modeling in \citet{Geoff_2021_J0924_AO_modeling} was used. However, as detailed lens modeling for B0445+123 and B0631+519 is not available, the Einstein radius was set to half of the quasar image separation measured from AO-based images.} Also, the source and deflector redshifts of these lens systems are known, except for B0445+123 and B0631+519, for which the source redshifts are still not measured. Thus, we have assumed a possible range of source redshift for these two lens systems. Using a flat $\Lambda$CDM cosmology with $H_0 = 70.0$ \hunit, $\Omega_{\rm m} = 0.3$, we have calculated the angular diameter distances and the corresponding SIS velocity dispersions of the deflectors. \newadd{In our sample, all the measured velocity dispersions in their mean are lower than the corresponding SIS velocity dispersions. These offsets are due to the vast simplification of the mass distribution assumption and chosen arbitrary cosmology. The deviation of the measurements from the respective SIS predictions is within 3\% - 20\%, except for B0445+123. The mean and scatter of the measured to SIS velocity dispersion ratios ($\sigma_{\rm measured}/ \sigma_{\rm SIS}$) are 0.87 and 0.08.} \referee{This systematic overestimation of the SIS predictions may not persist for a bigger sample size. We will explore this hypothesis in the future when such a sample is available. If this trend continues, it would be possible to estimate what modifications to the assumed galaxy mass model would minimize this trend.} The Einstein radii, source and deflector redshifts, and SIS velocity dispersion values are noted in Table \ref{table:veldis_measurements}. 
%In this regard, all the measured velocity dispersion values can be considered reliable. \\ 

\section{Conclusion}
\label{sec:conclusion}
The redshifts and velocity dispersions of the lensing galaxies are crucial data for measuring the Hubble constant using the time-delay strong lensing method. However, collecting stellar kinematics data with a high $S/N$ is challenging from seeing limited gound-based setups as the spectra of the lensing galaxies become spatially blended with those of the often much brighter lensed quasar images.  In this paper we have :
\begin{itemize}
    \item Developed a forward modeling technique and used it to extract 1D spectra of \newadd{eight} lensing galaxies, even in the presence of much brighter lensed quasar images.\\
    \item Used the extracted spectra to measure the deflector redshifts in all of our targets, including two that have never been measured before, in the J0818-2613 and J1817+2729 systems. For the remainder of the targets, our measured redshifts agree with those in the literature \newadd{to within 1$\sigma$}.\\
    \item Made the first measurements of the stellar velocity dispersions for six of \newadd{the eight lensing galaxies in our sample.} The new velocity dispersion measurements were made for the deflectors in  B0445+123, B0631+519, J0659+1629, J0818-2613, J0924+0219 and J1817+2729 systems.\\
    \item \newadd{Checked whether there exists significant covariance among the velocity dispersion measurements of the seven deflectors. In general, we found that covariances are negliglible except in a few cases, for which they should be taken into account.} \\ %we found that velocity dispersions do not vary in any preferential direction while fitted using the same setups.}\\
    \item Compared the measured velocity dispersions to those predicted by a SIS mass model \newadd{as a sanity check. In our sample, all the measured velocity dispersions in their mean are lower than the corresponding SIS predictions, and the deviations of the measurements are within 3\% - 20\% of the respective SIS velocity dispersions except B0445+123.} %, for which the deviation is bigger}.
    %, indicating that these lenses have total mass density profiles that are close to isothermal, $\rho(r) \propto r^{-2}$.
\end{itemize}
The Hubble tension is one of the major unanswered questions in current physics with immense consequences, and time-delay cosmography has proved its potential in resolving this issue. Though the \newadd{mass-sheet transform} (MST) poses significant uncertainty in the inferred \ho\ using this method, stellar kinematics such as single aperture velocity dispersion \newadd{can help} break this degeneracy \citep{TDCOSMO_IV}. %However, a larger sample \citep[$\sim$ 40 or more;][]{TDCOSMO_V} of time-delay lenses than currently published is required to achieve sub-percent precision.
\referee{However, to achieve percent-level precision, a sample of 40 or more time-delay lenses is required, for which we have measured integrated velocity dispersions of the deflectors to 5\% or better. Even in this scenario, kinematic measurements from a sample of $\sim 200$ nontime-delay lens systems are still necessary \citep{TDCOSMO_V}. The results in this paper will help to reach this percent-level precision by adding to the current sample.}
% a larger lens sample would minimize the uncertainties in the measurement process at the population level and help obtain a sub-percent precision. 
%The results in this paper will help to achieve such a sample size. 
%These results could also help in the modeling process of the respective lens systems. Overall, the results in this paper would pave the way to achieve a better measurement of the Hubble constant. \\

\begin{acknowledgements}
     %Discussion with these persons was helpful. 
     We thank Simon Birrer and Dominique Sluse for useful discussions and comments that improved this manuscript. PM and CDF acknowledge support for this work from the National Science Foundation under Grant No. AST-1907396. TT acknowledges support by the National Science Foundation through grants 1906976 and 1836016, by the National Aeronautics and Space Administration through grants HST-GO-15652 and JWST-GO-1794, and by the Gordon and Betty Moore Foundation through grant 8548. CS is supported by an `Hintze Fellow' at the Oxford Centre for Astrophysical Surveys, which is funded through generous support from the Hintze Family Charitable  Foundation.  \\ For the spectral resampling \textsc{SpectRes} \citep{SpectRes_2017} code was used. This research also made use of \textsc{numpy} \citep{Oliphant15}, \textsc{scipy} \citep{Jones01}, \textsc{Astropy} \citep{AstropyCollaboration13, AstropyCollaboration18},   \textsc{jupyter} \citep{Kluyver16}, \textsc{matplotlib} \citep{Hunter07} and \textsc{seaborn} \citep{Waskom14}. \\
      The data presented herein were obtained at the W. M. Keck Observatory, which is operated as a scientific partnership among the California Institute of Technology, the University of California and the National Aeronautics and Space Administration. The Observatory was made possible by the generous financial support of the W. M. Keck Foundation. The authors wish to recognize and acknowledge the very significant cultural role and reverence that the summit of Maunakea has always had within the indigenous Hawaiian community.  We are most fortunate to have the opportunity to conduct observations from this mountain. 

\end{acknowledgements}
\bibliographystyle{aa}
\bibliography{pritom_ref}

\end{document}